%% file: anonymize.tex
\documentclass[11pt]{article}

\usepackage{graphicx}
\usepackage{subfigure}

\begin{document}
\date{}
\title{\Large{\bf Sharing Computer Network Logs for Security and Privacy: A Motivation for New Methodologies of Anonymization}}
\author{Adam J Slagell and William Yurcik \\ National Center for
Supercomputing Applications\\ University of Illinois\\ \{slagell, byurcik\}@ncsa.uiuc.edu}
\maketitle

\input{abstract}
\input{ch1}
\input{ch2}
\input{ch3}
\input{ch4}

\input{ch5}

\input{ch6}

\end{document}

%% file: abstract.tex
\begin{abstract}
Logs are one of the most fundamental resources to any security professional. It is widely recognized by the government and industry that it is both beneficial and desirable to share logs for the purpose of
security research. However, the sharing is not happening or not to the degree or magnitude that is desired. Organizations are reluctant to share logs because of the risk of exposing sensitive information to potential attackers. We believe this reluctance remains high because current anonymization techniques are weak and one-size-fits-all---or better put,  one size tries to fit all. We must develop standards and make anonymization  available at varying levels, striking a balance between privacy and utility. Organizations have different needs and trust other organizations to different degrees. They must be able to map multiple anonymization levels with defined risks to the trust levels they share with (would-be) receivers. It is not until there are industry standards for multiple levels of anonymization that we will be able to  move forward and achieve the goal of widespread sharing of logs for security researchers.
\end{abstract} 

%% file: ch1.tex
\section{Introduction}
Log data is essential to security operation teams at any organization
large enough to have full-time security personnel. While IDSs operate
on streaming data, matching signatures and producing alerts, it is
still necessary for human beings to examine logs to understand these
alerts. Logs also form the core source of evidence for computer
forensic investigations following security incidents. The current
state-of-the-art is for each autonomous organization to use log data
to locally optimize network management and security protection. For
instance, it may only be when they themselves are scanned by an
individual that an organization will block a particular IP address.
Administrators may miss the bigger picture and not see that they are just a
piece of a larger target. Furthermore, administrators may only start
to scan their own network for a particular vulnerability once an
attacker has exploited it on their systems. There are very few
cross-sectional views of the Internet, and until recently there have
been no mechanisms to enable such wider views. Additionally, current examples
of wide views, such as spam blacklists and worm signatures, are often
focused on a specific characteristic even though signatures are 
gathered from events across the entire Internet. 

Sharing data is in fact common among attackers. They trade zombies, publicly
post information on vulnerable systems/networks and coordinate
attacks. Recent events at several U.S. supercomputing centers \cite{SuperAttacks} 
have demonstrated examples of coordinated attacks against
organizations that do not have good mechanisms of data sharing and log
correlation. Real, not simulated, data is necessary. While worms
that are let go without further human interaction could possibly be
modeled and simulated, human motives and specific interactions cannot.
It is no longer satisfactory to focus solely on the local picture; 
there is a need to look globally across the Internet. While the
data needed exists, tapping into thousands of data sources effectively and
sharing critical information intelligently and to the data owners'
satisfaction is an open problem. 

In fact, the Department of Homeland Security has recognized the
importance of sharing information and has established Information Sharing
and Analysis Centers (ISACs) to facilitate the storage and sharing of
information about security threats \cite{ISAC}. The importance of log
sharing has also gained industry recognition with investments in
infrastructure dedicated solely for this purpose across multiple
industry sectors \cite{NSSC}. The National Strategy to Secure
Cyberspace (NSSC) explicitly lists sharing as one of its highest
priorities---data sharing within the government, within industry
sectors and between the government and industry. In fact, of the
eight action items reached in the NSSC report, three of them are
directly related to log data sharing: Item 2:  ``Provide for the
development of tactical and strategic analysis of cyber attacks and
vulnerability assessments"; Item 3:  ``Encourage the development of a
private sector capability to share a synoptic view of the health of
cyberspace"; and Item 8:  ``Improve and enhance public/private
information sharing involving cyber-attacks, threats, and vulnerabilities". 

While it is understood and well-accepted that log sharing is
important, it happens on a very limited scale if at all. We believe 
that the problem, while social on the surface, is
technical at the heart. Organizations realize the importance of
sharing such data, but they are reluctant because sharing log data
allows their networks to be ``mapped out". This exposure creates an
increased risk for those who share. Anonymization of logs is still in
its infancy and the technology to date does not meet the needs of many
organizations. Even the largest public source of network traces
\cite{CAIDA} contains logs anonymized in inconsistent ways.  Some
data is anonymized, some is not. Anonymized data may preserve
prefixes or it may just truncate IP addresses. Even prefix-preserving
anonymization has different mappings between data collections from the
same sources but recorded at different times. Lastly, few data sets 
anonymize anything beyond IP addresses. 

We believe there is a need for standards in anonymization. There should
be defined levels of anonymization and methods to express those different
levels succinctly. There must also be ways to map an
organization's needs and trust levels with other organizations to the
appropriate anonymization levels. In addition to defining these
levels, methodologies and algorithms should be developed to
anonymize log data in new ways. 

The rest of this paper is organized as follows. Section 2 discusses
current efforts for sharing log data and related work in
anonymization. Section 3 covers the many types of logs and different
fields that could potentially be anonymized. Section 4 
discusses attacks against currently immature anonymization
systems. Section 5 discusses our goals and vision for a new system of
anonymization. We conclude in section 6.  

%% file: ch2.tex
 \section{Related Work}

\subsection{Current Internet Log and Data Collection Centers}
While there are a growing number of data centers that collect and
analyze logs, few are dedicated to sharing logs with the research
community. Those that do share do not anonymize sufficiently, and
they tend to focus on only one particular type of log. The goal is to
share as many types of logs from as many sources as possible.
Furthermore, there needs to be adequate protection of these logs to suit
the different needs of those who share them. To accomplish this goal,
there must be multiple, standardized levels of anonymization. 

\subsubsection{CAIDA}
CAIDA \cite{CAIDA} has one of the largest Internet log data catalogs
available for public analysis. They have an index to many data
sets stored off site and are developing a cataloging system for these
data sets. They also seek to develop tools to analyze the data sets.
CAIDA focuses on macro-level data to support network measurement
research. As such, they do not have the level of detail or the many
heterogeneous types of logs necessary for security research. The
usefulness of these logs to security research is limited to a very high level. 
The network traces that CAIDA provides could be used to get a big
picture of worm activity, but they do not contain the level of detail to
capture new exploit binaries or hacker behaviors post-compromise. 

Even though these logs are not primarily intended for security
research, they still suffer from the problems of undefined
anonymization standards. While it is important to anonymize logs so
that adversaries cannot map out contributor networks, CAIDA itself
does not anonymize logs that it does not generate. 
Different organizations anonymize to different levels---some not
anonymizing at all---and in different manners. There is no
consistency in the way anonymization is done because all of the
contributors do it their own way. Thus we see that development of
anonymization standards applies to more than just security research. 

\subsubsection{DeepSight}
Symantec's DeepSight \cite{Symantec} does utilize a cross-sectional view of the
Internet and analyzes detailed IDS, firewall and virus scanner logs.
However, it is not a log sharing system for research, but rather it is
a data collection system with a commercial purpose. Anonymization is 
not a real issue because data is not shared, but rather all data is 
collected and sent to a trusted third party. The purpose of this
collection is to notice trends and provide early warnings of attacks
and new threats to customers. They allow certain thresholds and other
variables to be set by customers to somewhat customize their alerts and 
reports, but mostly it is a way to alert their customers of
new worms, viruses or software exploits being used in the wild. It
does not provide data sets to share with the research community. 

\subsubsection{Internet Storm Center}
The Internet Storm Center (ISC) \cite{SANS} is like a grass-roots
version of DeepSight that is run by SANS. They collect IDS logs
from volunteers and analyze them to detect trends. Their purpose is
to provide an early warning system of new worm activity on the
Internet. They provide reports on the top ports being scanned with
respect to time, and they use the trend information they find to
determine the INFOCon threat level, much like Symantec defines the
ThreatCon level with DeepSight data. 

ISC does not share actual logs, but they produce high level statistics.
For this reason, their port activity and trends data do not need to be
sanitized. They sanitize information about scanner source IPs by looking very
broadly at the number of scans per class C network. This kind of
anonymization is also used in some of the CAIDA logs where they simply
truncate IP addresses. The danger is pretty low in sharing this kind
of information, but its usefulness is also minimal. It does nothing
more than allow inferences such as ``The US does the most scanning" or
``Universities contribute to most of the P2P traffic". Many of these
statistics can 
be predicted from the density of addresses assigned in
the respective class C networks. ISC does share specific addresses in
one place: it lists the top 10 scanners by IP address. Many
organizations use this information to block misbehaving machines. The
repercussion of doing this is again minimal. They do not provide
specific details about those machines or the networks they are on. It
simply serves to embarrass the ISPs that host the compromised
machines. Any sort of anonymization here would defeat the purpose. 

Overall, the type of data they provide is a homogeneous set of aggregated  
statistics. More information can be gathered from CAIDA logs because 
raw access is provided. Thus, one is not restricted to only the
statistics they provide. The main difference of course is that the ISC
is real-time and uses  
a more distributed sample. In conclusion, ISC works very well for monitoring
general worm behavior, detecting trends that indicate new worms and
analyzing the life cycle of an exploit. However, they are not
gathering many types of logs, and they are not sharing them with the
general community for research. 

\subsubsection{DShield.org}
DShield.org \cite{DShield} is a grass-roots log collection system,
though it is now 
funded partially by SANS. They gather firewall logs and convert them
to a standard format. Currently, they exclusively accept packet filter
traces. These are used to create reports of types similar to the
Internet Storm Center. They have reports on port activity trends, the top
10 most offensive scanners and the top 10 most probed ports. They
produce the blacklist of offenders that the ISC uses and provide
searches on activity by particular IP addresses. 

Anonymity is not dealt with seriously here, and they say ``You should not
submit any information you consider business critical or proprietary".
They say that they ``try" to hide destination IPs to mask who is being
attacked, but raw data is searchable and may be made available raw
to the public. Of course they do not indicate the submitter of the
data. Decisions of what and how to release raw data is made on a per
individual basis. 

Several problems exist with the system as is. First, there is only
one type of log. Second, there are no precautions to keep people from
resubmitting logs and polluting the data set. If special clients are
used instead of web submissions, accidental resubmissions are prevented. 
Third, anonymous submissions allow fake data to be
submitted that could wrongly blacklist individuals. Fourth, even if
their non-guaranteed anonymization of target hosts works, anyone can
query information about specific hosts and networks. This allows
attackers to find already compromised machines on a network rather
easily. In conclusion, DShield.org provides data of limited types and
minimal data protection mechanisms. 

\subsubsection{Packet Vault}
The University of Michigan has worked on a secure, long-term archive
of network packet data they call the {\it Packet Vault} \cite{Antonelli00, Antonelli99}. 
It is basically a special purpose network device and encrypted database system 
specifically designed for packet sniffers. It is designed so that selected
traffic can be made available without exposing other traffic.

They use the ``black marker" approach to anonymize logs by completely
encrypting packet information. As such, it cannot really be called
anonymization because all fields are either encrypted or decrypted,
essentially having the same effect as printing a log and using a black
marker on most of the lines. They group items under the same
encryption key if the packets are part of the same ``conversation". 

Now, if instead they used different keys for different fields, that
would allow them to release different views of the same records to
different organizations. That would be a crude sort of black marker
(all-or-nothing) anonymization of selected fields. But their goals
are different. They are not trying to share logs while preserving
privacy. They are making logs available to participants of
the conversation, but not to anyone else. They give the
appropriate keys to decrypt logs records that describe a participant's 
actions but not those in which they were not a participant. 

In conclusion, we see that while there are some centers dedicated  to
collecting log files, they all suffer from one or more of the
following problems: (1) They do not have a wide view of the Internet
but are quite localized, (2) the repositories are very specific,
addressing one or only a few types of logs, (3) anonymization is weak
or nonexistent and usually inconsistent or (4) they collect many logs
but do not share them with the research community. 

\subsection{Anonymizers}
There has been some research in log anonymization. However, most
work addresses only a small subset of all the available log
sources (particularly network traces) and focuses exclusively on anonymizing
IP addresses within a log. While IP addresses could simply be removed 
or randomized in logs,  such a solution is undesirable since it
destroys a basic structure used in analyzing logs. Significant work
has been accomplished on prefix-preserving anonymization of IP 
addresses. In prefix-preserving anonymization, IP addresses are mapped
to pseudo-random anonymized IP addresses by a function we will call
$\tau$. Let $P_n()$ be the function that truncates an IP address to $n$
bits. Then $\tau$ is a {\it prefix preserving} permutation of IP addresses
if $\forall$ $1\leq n\leq 32$, $P_n(x)=P_n(y)$ if and only if
$\tau(P_n(x))=\tau(P_n(y))$. TCPdpriv \cite{TCPdpriv} is a free program 
that performs prefix-preserving TCPdump trace anonymization using
tables. Because of the use of tables, it is difficult to process logs
in parallel with this tool. In \cite{Xu01,Xu02}, Xu et al.\ have
created a prefix-preserving IP pseudonymizer that overcomes this limitation by
eliminating the need for centralized tables to be shared and edited by
multiple entities. Instead, with their tool CryptoPAn, one only needs
to distribute a short key between entities that wish to pseudonymize
consistently with each other. Furthermore, they have shown that all
prefix-preserving pseudonymizers must take a particular form and that
their solution is optimal with respect to security. But while
theoretically optimal solutions have been created for this reduced
problem, the larger problem of anonymizing whole log files 
remains unsolved. 

One of the earliest uses of {\it pseudonyms} can be found in
\cite{Chaum81} where public keys are used as pseudonyms. We now
recognize what Chaum described in \cite{Chaum81} as a ``digital
pseudonym" to be a specific type of pseudonym called an authorization
certificate. As noted in \cite{Lundin99}, pseudonyms help define
middle ground in the zero-sum tradeoff between security and privacy of
audit logs. In \cite{Sobirey97}, Sobirey et al.\ first suggested
privacy-enhanced intrusion detection using pseudonyms and provided the
motivation for the work of Biskup et al.\ in \cite{Biskup00a,Biskup00b}.

While the work in \cite{Lundin99,Biskup00a,Biskup00b} does deal with
log data and anonymization, their goals are significantly different
than ours. All three works deal specifically with pseudonymization
in Intrusion Detection Systems (IDSs). The adversary in their model
is the system administrator, and the one requiring protection is the
user of the system. In our case, we instead assume that the
system/network administrators have access to raw logs, and we are
trying to protect the systems from those who would see the shared
logs. To contrast how this makes a difference, consider that in their
scenario the server addresses and services running are not even
sensitive---just information that could identify clients of the
system. Furthermore, we do not care about reversal of pseudonyms. We
have no need for that capability, but since the system/network
administrators do not have raw data in their case, the privacy officer
must help the system security officer reverse pseudonyms if alerts
indicate suspicious behavior. In \cite{Biskup00a,Biskup00b}, they
take this further and try to support automatic re-identification if a
certain threshold of events is met. In that way their pseudonymizer
must be intelligent, like an IDS, predicting when re-identification
may be necessary and thus altering how it pseudonymizes data. They
also differ from us in that they create transactional pseudonyms,  
so a pseudonym this week might map to a different
entity the next week. We, however, require consistency with respect to time for
logs to be useful. Lastly, all of the anonymizing solutions in these
papers  filter log entries and remove them if they are not relevant to
the IDS; we dispose of no entries because completeness is very important
for logs released to the general research populace. 

In \cite{Flegel02}, Flegel takes his previous work in privacy
preserving intrusion detection \cite{Biskup00a,Biskup00b}  and
changes the motivation slightly. Here he imagines a scenario of web
servers volunteering to protect the privacy of visitors from
themselves, and he believes IP addresses of visitors need
pseudonymization. However, to a web server IP addresses already act
as a pseudonym protecting the client's identity since ISPs rarely
volunteer IP address to person mappings. The case where this is not
true is if the web server is that of the ISP. Then in that
particular instance IP addresses are not pseudonyms. Though the
motivation differs slightly, the system described is the same
underlying threshold based pseudonymization system, and the focus of
this paper is really about the implementation and performance of the
system. As such, the results of \cite{Flegel02} can be applied to
\cite{Biskup00a,Biskup00b}. 

In \cite{Pang03}, Pang et al.\ developed a new packet anonymizer that
anonymizes 
packet payloads as well as transactional information, though their
methodology only works with application level protocols that their
anonymizer understands (HTTP, FTP, Finger, Ident and SMTP). The
process can also alter logs significantly, losing fragmentation
information, the size and number of packets and information about
retransmissions, skewing time stamps, sequence numbers and
checksums. 
While their anonymizer is limited in its capabilities, it is fail-safe
because it only leaves information in the packets that it can parse
and understand. Further, they create a classification of
anonymization techniques and a classification of attacks against
anonymization that we found useful. We use a similar classification
which is based off of their work. 

Most recently Waters et al.\ \cite{Waters04} address the tension
between data access control and searchability of audit logs through a
new method they developed to search asymmetrically encrypted logs. In
this way the encrypted log can be made public for search, and the
owner distributes private keys corresponding to keywords. Thus,
instead of the data owner decrypting the log and running the search,
he can simply give the query maker the ability to perform the query
with a set of keywords he deems acceptable.

%% file: ch3.tex
\section{Log Varieties}
A computer network contains a variety of different infrastructure
devices, each of which may be instrumented to produce multiple audit
logs. Although the topic of computer and network audit logs is broad, a
topic of its own, we feel a brief survey of some of the different
types of logs is an important starting point in understanding the
issues of sharing heterogeneous logs for network measurement and
security research.

Note that we are making it a point to emphasize {\em heterogeneous}
audit logs. The fact that the audit logs are different is significant
because it promotes multiple views for discovery, robustness against
attack, interoperability, extensibility and flexibility. However,
heterogeneity also provides new avenues of attack against the
anonymization system. While one type of log may not be enough to
break the anonymization system, information may be inferred from
multiple logs that can be used in a successful attack against
anonymized data. Thus we seek to create anonymization schemes for
many types of logs.

What follows is a description of commonly implemented network and system logs
summarized from \cite{Yurcik03}. These logs provide situational
awareness of what is happening where and when on networks/systems by
auditing system activities, transactions performed and network
signaling. These logs are useful for detecting network problems,
malicious activity and recovery from accidental or intentional
failures. 

\subsection{TCPdump}
One of the most common ways of collecting network data into
logs is through use of the TCPdump utility. This utility captures
headers from packets on a network interface set in promiscuous mode
and displays the binary data in a human-readable formats. 

While TCPdump is a valuable tool, it focuses only on the TCP/IP suite
of protocols. There are a variety of other utilities for sniffing raw
packets of any protocol from any point on a network. Referred to as
{\it sniffers}, the most common examples are the open source tool 
Ethereal and Sniffer from Network General Corporation---until 
recently Sniffer was owned by McAfee. As networks
increasingly employ switch technology, sniffers that rely on a shared
medium network (e.g. traditional Ethernet) are being moved from end
systems to servers and routers (and now wireless networks).
Alternatively, commercial switches often employ special ports
created particularly for sniffers to tap into. Sniffer logs are
valuable in discerning low-level attacks such as abnormal traffic
attacks (e.g. 802.11 ARP poisoning); however, their scope is limited
by their monitoring position within a network, and the log size makes them
cumbersome to analyze. 
 
\subsection{NetFlows}
NetFlow logs contain records of unidirectional flows between
computer/port pairs across an instrumentation point (e.g. router) on
a network. Ideally, there is an entry per socket. These records can
be exported from routers or software such as ARGUS or NTOP. NetFlows
are a rich source of information for traffic analysis consisting of
some or all of the following fields depending on version and configuration:
IP address pairs (source/destination), port pairs
(source/destination), protocol (TCP/UDP), packets per second, 
time-stamps (start/end and/or duration) and byte counts.

\subsection{Syslog} 
Syslogs are a UNIX standard for capturing information about networked
devices, daemon processes and even kernel messages. Messages are
encoded by level (e.g. warning, error, emergencies) and by facility
(e.g. service areas such as printing, e-mail and network). Syslog can also 
serve as a distributed error manager by forwarding log entries to
centralized syslog servers for processing. Syslogs can be
pattern-matched for known attack signatures. They can also be
searched for potentially suspicious activities such as critical events
(e.g. modules being loaded and core dumps), unsuccessful login
attempts, new account creation (especially accounts with special
privileges), suspicious connections to unused ports, or simply the
cessation of logging messages from a host ( which may indicate the logging
process has suspended or logs wiped).

\subsection{Workstation Logs}
Workstation logs are standard utilities that keep login/logout entries
on a workstation's local hard disk (e.g. Window's event viewer). Some
application software also maintain access logs. For example, virus
scanners maintain logs of previous scans. Virus scanners and mail
agents themselves may log all outgoing mail messages as
well. Workstation logs are enabled by default on most operating
systems, but it is almost always possible for an adversary with 
escalated privileges to disable them.

\subsection{ARP Cache}
Routers and switches contain cached tables of recent resolutions of
MAC addresses to IP addresses called ARP (Address Resolution Protocol)
caches. The entries are of two types: {\em dynamic} entries that are
added/removed automatically over time and {\em static} entries which
remain in the cache until the computer is restarted. Each dynamic 
ARP cache entry has a potential lifetime of between 2 and 10 minutes
(depending on operating system settings, traffic levels and cache
size),  and a log of all entries can be created over a specified time
period. 

The ARP cache is useful for determining static IP addresses, 
identifying unregistered/unknown (including maliciously spoofed) and
misconfigured devices attached to a network, detecting certain layer 2
attacks (e.g. ARP poisoning), identifying what IP address(es) a
particular hardware address is using, to debugging connectivity problems
a device may be experiencing and tracking unsuccessful connection attempts to
devices that either are not currently on the network or do not exist
(e.g. port scans to non-existent hosts). These logs are
becoming more important with the recent growth of wireless networks
and ease at which it is possible to perpetrate ARP poisoning and
man-in-the-middle attacks against them. 

\subsection{DNS Cache}
DNS (Domain Name Server) caches contain mappings between
fully-qualified hostnames and their corresponding IP addresses based
on recent requests to other name servers. The amount of time a name
server retains cached data is controlled by the time-to-live (TTL)
field for the data. These logs can be created via periodic (period is
shorter than minimum TTL) snap shots of the cache timed to capture
data at least once before it expires. Host tables (.rhosts and
hosts.equiv), which also map hostnames to IP addresses, provide static
mapping information. DNS cache records provide useful evidence
of attacks purported against DNS services and sometimes of DDoS 
worms that create high volumes of DNS queries for a target.
 
\subsection{Dial-up Servers}
Dial-up server logs maintain system accounting records of who makes
incoming network connections. They are a very reliable source of
information to investigators because the log is difficult to poison
with false information. Even if an attacker steals another's
credential to login, the telephone records are very difficult to fake,
thus giving away the attackers location if nothing else. However, as
VoIP becomes more prevalent, the reliability of telephone records may
diminish somewhat. Also, dial-up connections are much slower than
many attackers can tolerate, and with the prevalence of broadband, attacks
through dial-up servers have become less common. We expect VPN logs
to replace the role of dial-up server logs in the near future.

\subsection{Kerberos}
Kerberos logs contain records of all ticket requests and
uses. This information can be used to generate login graphs and
determine who was logged into a particular workstation at a particular
time. This may help in detecting tickets that have been compromised,
perhaps by a brute force password attack, and used by automated tools
and scripts.

\subsection{SNMP} 
SNMP (Simple Network Management Protocol) logs, referred to as
Management Information Bases (MIBs), are databases of managed objects
that store information about a wide variety of network devices. 
The SNMP operator application monitors network devices via
polls to network device agents for specified MIB information or traps
from network device agents notifying the operator of an event.
 
\subsection{Routing Tables}
Routing table logs (e.g. inter-domain BGP, intra-domain OSPF or RIP)
provide information about routing-based attacks and errors ranging
from individual misbehaving routers that drop/misroute packets or
inject disruptively large routing tables to the systemic
network-wide advertisement of false routing information or instability
caused by the propagation of worms. Global, local or peer routing
tables all provide different vantage points for analysis.
 
\subsection{Firewalls}
A firewall is a computer or network device that interfaces between an internal
network/computer and external networks that are trusted to a lesser
extent (e.g. the Internet) to enforce an organizational access control
policy by processing packets/connections based on the rule set. Note
this definition is being expanded as there are now personal firewalls
that are being installed on workstations. These differ in that they
are positioned at the endpoints and hence can be more application aware. 
Firewall logs are important in a recursive way, to maintain the
effectiveness of the firewall's internal rule set. A rule set exactly specifies
what traffic to permit/block and typically grows in the number of
rules beyond human comprehension in a commercial setup. 

Firewalls can be used to monitor both normal activity (types of
services requested and used, common external IP addresses accessing
internal services, common access time patterns) and suspicious
activity (probes to ports with no authorized services,
external-to-internal flows with spoofed internal IP addresses,
out-bound connections from uncharacteristic internal machines and
modification/disabling of the firewall rule set).

\subsection{Intrusion Detection Systems} 
Logs from Intrusion Detection Systems (IDS) contain alerts indicating
specific attacks that have occurred. Generally, while a firewall has a
proactive, preventative focus, an IDS has a passive, reactive focus.
The assumption is that eventually some one will break through the
perimeter firewalls, and then one needs to be able to detect
intruders. IDSs can be categorized by sensor
placement (network versus host) and by technique (signature versus
anomaly), with all types producing both alerts and detailed logs.
Real-time IDSs have been plagued by large log sizes and high false
positive rates---especially in anomaly based systems, but incremental
improvements are increasing their effectiveness for post-mortem forensics.
 
\subsection{Mail Servers}
Mail logs maintain a log of completed transactions (as well as a queue
of pending transactions) including the sender and recipient addresses,
subject titles, time-stamps and file sizes. Common log reports
generated include: total length of time spent receiving and sending
e-mail, the number of e-mails by an entity (organization, group, or
individual) over a specific period of time (day/week/month),
stratification of e-mail by time (work hours/off-hours) and common
addresses, stratification by size and type of file attachments and
identification of dormant accounts.
 
\subsection{Web Servers}
Web server logs have traditionally been used to provide feedback on
performance and misconfigurations (e.g. link errors). Web logs provide
detailed records of requests to the web-server and statistical
information about network traffic. Web log record attributes can
include: the source IP address from which a request was generated,
whether the request was satisfied, a userid determined by the HTTP
authentication, a status code and the size of the object returned
with each satisfied request.

While traditionally used for performance information, web logs are
being used more frequently for security analysis. They can be used
to detect illegitimate requests (e.g. asking to run a script in a
directory that should not be accessible) that exploit
misconfigurations, buffer overflow attacks on the web server used to
run arbitrary processes with the privileges of the web server daemon
and attacks targeting specific web applications and scripts that are
not secure.

\subsection{DHCP}
Dynamic Host Configuration Protocol (DHCP) server logs can be used to
track IP address assignments to devices as they join/leave a network.
DHCP servers manage two databases: (1) an Address Pool database for
holding IP addresses and other network configurations and (2) a
Binding database for mappings between hardware MAC addresses and an
entry in the Address Pool. Though most frequently used to assign
dynamic IP addresses, DHCP can also assign a dedicated static address
for a device that re-joins. On a network that uses DHCP with dynamic
addresses, maintaining a log is absolutely necessary to be able to
forensically associate dynamically changing IP addresses to specific devices.

\subsection{Scanners}
Scanners for defensive purposes are used to perform risk management by
detecting vulnerabilities and notifying system administrators of the
vulnerabilities and patches that need to be installed. They typically
generate reports rather than logs, though. Scanners range from simple
port scanners such as NMAP that report open ports and operating
systems detected, to very advanced scanners such as NESSUS that runs NMAP to
detect open ports and determines exact versions of services running, and
whether these systems are vulnerable to known exploits. Along with
the rise in managed security, there are companies such as Qualys that
provide proprietary scanners through a web interface and scan for
you. Qualys, for example, will scan from the outside---or inside if
you purchase a special network device---and produce a complete,
customizable report in multiple formats that indicates what
vulnerabilities your systems have. With these managed services you
typically get easier to read reports and the most up-to-date
vulnerability databases.

While this list of sixteen log types may seem exhaustive, it is not.
The list of potential log sources are more numerous than the number of
services and daemon processes in deployment.  There are many
proprietary logs  we did not mention, including reference monitor
alerts, router traps and a myriad of application software
logs---though many of the latter log through syslog on UNIX based
platforms.  The challenge to those working on the problem of log
anonymization for security is that they must consider as many log
sources as possible and try to generate a basis of logs that contain
as much information as possible with minimal over lap between logs.  

%% file: ch4.tex
\section{Attacks Against Current Anonymization Schemes}

Anonymization---and particularly pseudonymization---schemes are
difficult to secure. An anonymization scheme is said to be secure if
one cannot link records or information to specific entities
(e.g. hosts or users). Anonymization schemes try to achieve balance
between security and utility---utility being a measure of how much
useful information remains after anonymization. Schemes that can be
proven secure tend not to be useful because of severe information
loss. Pseudonymization tends to have more utility, but there are often
more attacks against such schemes. In what follows we examine five
classes of attacks on anonymization and pseudonymization
schemes. These are similar to what Pang et al.\ define in
\cite{Pang03}. These attack types are the basic building blocks of
more advanced attacks and are often combined together and used against
multiple heterogeneous logs. 

\subsection{Fingerprinting}
We closely match Pang et al.\ by defining fingerprinting as {\it the
  process of matching attributes of an anonymized object against
  attributes of a known object to discover a mapping between
  anonymized and unanonymized objects}.  For example, consider an
anonymized NetFlow log of an organization with only one web server
with an IP address of 192.168.77.29. Suppose that we see an anonymized
IP address of 10.19.21.3 that is responding to port 80
connections. Further, 95 per cent of all port 80 connections are going
to this address. Then we can infer that 192.168.77.29 maps to
10.19.21.3. This is most likely correct unless someone is running an
illegal web server with a tremendous amount of traffic. Fingerprinting
is most useful in identifying servers because of the unique attributes
they have. 

\subsection{Structure Recognition}
Structure recognition is {\it the act of recognizing structure between
  objects to use one mapping to discover multiple mappings between
  anonymized and unanonymized objects}. By themselves these attacks
reveal no mappings, but used with a known mapping they can reveal new
mappings. One example would be a common attack against all
prefix-preserving IP address pseudonymization schemes. Knowing one IP
address mapping in this case reveals bits of any addresses matching
prefixes with the address of the known mapping. Here we are exploiting
the structure of CIDR addressing and the structure-preserving
properties of the anonymization technique. Another example could be
the recognition of a port scan in an anonymized trace. This could
reveal a sequential ordering of anonymized addresses. Knowing just one
mapping of anonymized to unanonymized addresses would reveal all
mappings for the scanned machines. 

Notice that our definition is more refined than Pang et al.'s. Their
definition is ambiguous enough to allow overlap between fingerprinting
and structure recognition. For example, one could recognize the
structure of traffic patterns to determine a gateway server. By their
definition that could be fingerprinting or structure recognition. We,
however, do not count that as structure recognition. Structure
recognition itself does not reveal mappings. Rather, it discoveries
relationships that allow the discovery of one mapping to aid in the
discovery of more mappings. 

\subsection{Known Mapping Attacks}
Known mapping attacks {\it exploit the discovery of a mapping between
  unanonymized and anonymized data in one reference, to undo
  anonymization of that same data in multiple places}. This kind of
attack can occur in two manners. For example, a username and password
field may be anonymized in exactly the same manner. In that case the
word ``Emily" could be used as a username in one record and a password
in another. Revealing the mapping in one field would reveal it in
another. A similar attack could be carried out if IP mappings are
consistent across multiple logs. In that case, revealing a mapping in
one log reveals the same mapping in another. 

\subsection{Data Injection}
Data injection attacks {\it involve an adversary injecting information to be 
logged with the purpose of later recognizing that data in an anonymized 
form}. To illustrate this type of attack, imagine the following scenario. 
Your organization has wanted to help the research community and has 
just heard about prefix-preserving anonymization. They now feel safe 
to share logs and decide to publicly post anonymized logs on a regular 
basis. An adversary, Anton, knows that anonymized logs are regularly 
posted from your organization and like everyone else has access to
them.  Anton, knows that if he had an internal view of your
network, he could see a lot more about how machines interact and are
setup. He can send limited probes to certain machines, but a lot of
them are filtered and his IP address is often quickly blocked. He
knows the logs from the firewalls, routers and IDS would help a lot,
but they are protected by prefix-preserving anonymization. However,
Anton quickly realizes that if he could recognize his own scans in the 
anonymized logs, he could figure out the mappings of several IP 
addresses through a structure recognition attack.  Because of
the prefix-preserving property, every mapping he discovers yields many
bits in the IP addresses of machines he did not even scan. So with just
a few scans, he can glean information about almost all of the
pseudonyms. The only problem left is how does he make scans that he
can later recognize. Well, this is easier than setting up a covert
channel. His scans need only make use of special sequences that are
not random, but could look random without tenacious investigation.  

A quick solution might be to use something as simple as the
Fibonacci sequence. A smarter solution would be to use a sequence
determined by PRNG seeded with a password. For instance, what if Anton
scanned a particular machine with the target or source ports being
Fibonacci numbers?  That should be very recognizable in a log if you
know to look for it. He could do the same thing with timings
between probes. Of course, the timing intervals must be large enough 
that jitter in the network does not  cause mismatches
on a pattern. In fact, such an attack could be made with a number of
fields, even in obscure TCP options. 

\subsection{Cryptographic Attacks}
Many anonymization techniques depend on cryptographic functions. Such
systems can be vulnerable to attacks on the cryptographic algorithms,
such as known or chosen plaintext attacks. These attacks could reveal
secret keys or other information. Such attacks don't affect just
single objects, but all entries of a particular field in a
log. However, cryptography is less likely the weak link in a
scheme. It is usually easier to purport the other types of
attacks. Also, this kind of attack typically requires mappings to
first be discovered through other attacks. Of particular interest is a
data injection attack which could then be used to attempt a chosen
plaintext attack against the cryptographic algorithms.

%% file: ch5.tex
 \section{Goals and Approach}
Our grandest goal is to standardize anonymization techniques across
most log types through IETF drafts and RFCs. We would also like to
develop prototype anonymization tools and document scenarios of
mapping trust levels between organizations to the newly standardized
anonymization levels. However, many steps must be taken to reach
these goals.

\subsection{Requirements Gathering}
In defining different anonymization levels, we begin by defining what
we mean by anonymization. Anonymity is the state of not being
identifiable within a set of potential subjects. In communication
scenarios, both senders and receivers can be anonymous. Pseudonymity
is a special type of anonymity and is often what we mean when speaking of
anonymity. Specifically, providing pseudonymity is the ability to
prove a consistent identity without revealing one's actual
identity by use of an alias or pseudonym. Simply encrypting the
identity would be an example of pseudonymity because we would still
have unique identifiers. However, masking or removing identity
fields is not providing pseudonymity since a particular subject's set
of activities cannot be identified. Ideal pseudonymity only has
benefits as it means to be anonymous and able to separate the behavior of
unidentified individuals. As such, less data is lost, and
the same protection is provided. However, ideal pseudonymity is not
always obtainable, and it is difficult to prove a system has that property. 
Often one finds certain attacks may be more effective
against pseudonyms than systems that simply erase all identifiers. As
such we will be using both anonymity and pseudonymity with the latter
being preferred.

\subsubsection{Log Identification and Correlation}
We have already begun a very important first step in requirements
gathering: we have been surveying the many types of logs available.
But by no means have we exhaustively listed all log types in Section 3. Other
useful, proprietary logs include reference monitor logs, router traps
and a myriad of application  software logs---though many UNIX
applications log through syslog. We must eventually narrow this list
of logs to those most useful to security researchers. 

It is also important to be able to classify the logs by fields and
type of information. We 
will most likely find that similar logs have similar anonymization
solutions, but it is a real challenge to combine information from all
these logs in a useful way. While often the attributes initially appear
redundant---since they are contained in multiple logs---one finds that
this overlap is vital to enable event correlation between logs. 

Given this set of potential log files, the more logs that can be
processed for attribute correlation the better. However, the dominant
constraint, computational power, limits the number of log files which
can be quickly analyzed in real-time. In fact, storage can even become
a limiting factor when analyzing large network traces from busy
routers. Most Juniper routers cannot even make complete traces, but instead
sample the data. In \cite{Yin03}, Yin et al.\  discuss how to find a subset of
logs that are orthogonal---with each log providing some
information that is not present in the other logs---as well as
complete. It is ideal to find such a reduced set that provides
maximal information about network/system status. In mathematics 
such a set would be called a basis. Given this goal,
\cite{Yin03} initially divides the entire set of potential logs into two
categories: those logs that provide information about systems and
those logs that provide information about the network. It is likely
that other equally valid ways to divide the entire set of potential
logs can be used, but we have found it useful to first make this
separation in  categorizing logs since there is usually minimal overlap
between network and system logs. On the other hand, we often
find that when comparing two network logs, one may almost be
just a subset of the other log.

\subsubsection{Putting on the Black Hat}
We and others have thought of a few types of attacks against
anonymized logs, as discussed in the previous chapter. However, we
have just begun. Once the types of logs that are crucial to share are
identified, we must put on our black hats and think about how an
attacker can glean privileged information from them. This includes
identifying what is privileged and sensitive information in the
beginning, and it means putting back on our black hats when evaluating
the mechanisms we have developed to protect logs. As mentioned,
attacks against current anonymization systems are simple, effective
and fairly well known. That is in fact one of the main reasons
parties are still reluctant to share logs. There are no good,
holistic solutions as of yet. So, we must not only create solutions
resilient to the current attack methodologies, but we must
also look for potentially new attacks. 

\subsection{Defining Standards and Beyond}
Obviously different organizations trust each other for different tasks
and to different levels.  Anonymization is a balance between security,
meaning it is difficult to break the anonymization scheme or exploit
the protected information, and utility, meaning the usefulness of
the information to other parties. For example, in the previous
chapter we saw that prefix-preserved IP addresses do not protect
NetFlows from an active adversary. Anyone who can inject traffic to
be logged can begin to determine actual IP addresses. A solution
may be simple truncation or deletion of the IP address fields, but
now pseudonymity is lost. One attacker cannot necessarily be
distinguished from another. Still, port numbers can give away
information, particularly if used with time-stamps. One could add
noise to the time-stamps and delete port numbers, but now what useful
information is left in the logs?  

Organizations must determine how
much they trust organizations that may see their logs and think of
them as an adversary to some extent---by this we mean they must
determine what kind of attacks are likely from parties who will see
their logs. This, combined with the need-to-know level of the
receiving party, will determine how to anonymize the logs. One may
say this is easy because you simply never give more information out
than is necessary, but our response to this is that you often do not
realize how much information you are giving out. 

Ideally, we would like to develop anonymization levels that are
completely independent of the specific log type. Instead, they would
depend upon the fields in the logs---of which there are many fewer
types than of logs themselves. Even if we achieve such a goal, logs
are still in many different, incompatible formats. We would thus like
to create prototype tools to anonymize the most common types of logs.
We also intend to develop a standard XML format to which we can output
anonymized logs. Of course, it is always important to have the
ability to anonymize logs while retaining the original format as well.
This enables the same tools to operate on anonymized and unanonymized
logs. 

While our main goal is to define standards and our secondary goal is
to develop tools, there is one more important component. People must
understand how to apply the standards. Having different levels of
anonymization is not enough. People must be able to map their
specific situations and needs to the appropriate anonymization level.
This is in effect a need to map trust levels to anonymization
standards. For example, as a university we may trust graduate student
researchers more than some lab in a foreign country for the simple fact that we
can take action against students who exploit the privileged information. Plus,
those with campus IP addresses already have a better view of our internal
network. Thus, a lower level of anonymization may be needed for
sharing logs with students. We intend to create examples and
scenarios to help implementors use the new anonymization technology. 

\subsection{Architecture}

\begin{figure}
\centering
\includegraphics[scale=0.4]{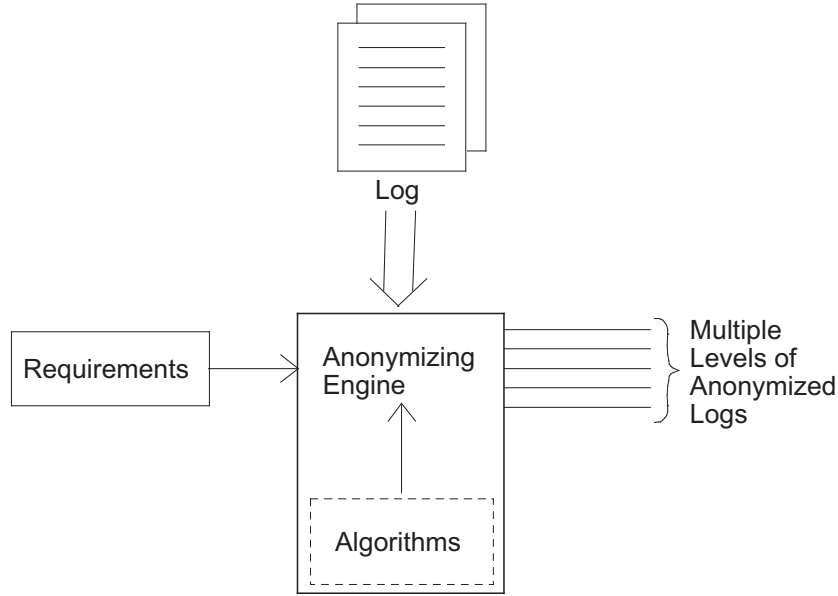}
\caption{\footnotesize {\bf Anonymizer Architecture}}
\end{figure}

Figure 1 presents the high-level architecture of the new anonymization
system. Note that this system accepts heterogeneous network logs as
input along with requirements profiling the level of anonymization.
It is anticipated that different encryption algorithms will be needed
for different types of data fields, so the encryption algorithm is a
component plug-in to the anonymization engine. The output will be log
files at varied levels of anonymization corresponding to the input
requirements profile.

%% file: ch6.tex
 \section{Conclusion}
Logs are vital to both security operations and researchers. Security
professionals analyze them on a daily basis. Both government
institutes and industry have recognized the importance of logs and
furthermore the value of sharing logs with the security research community. Even
though the importance of sharing logs is widely recognized, it is not
happening on a large scale. While there are centers dedicated to
collecting logs, they suffer from one or more of the following problems:
(1) They do not have a wide view of the Internet but are quite
localized, (2) the repositories are very specific, addressing one or
only a few types of logs,  (3) anonymization is weak or non-existent and
usually inconsistent or (4) they collect many logs but do not share them 
with the research community.

We believe the problem, while social on the surface, is technical at
the heart. Organizations realize the sensitivity of their information
and that current anonymization techniques are inadequate and
immature. Hence they are reluctant to share them because of the
risk. We contend that the solution is standards and new methods of
anonymization. There must be multiple levels of anonymization for the
many and differing needs of organizations. Organizations need to have
choices that can be clearly described and associated with specific
risks so that they can choose the appropriate anonymization
level corresponding with the trust level they have with those to whom
they would share logs. It is only then that this reluctance to share
logs will be overcome.